\documentclass{Interspeech2024}

\interspeechcameraready

\usepackage{subcaption}
\title{Natural language guidance of high-fidelity text-to-speech with synthetic annotations}
\name[affiliation={1}]{Dan}{Lyth}
\name[affiliation={2}]{Simon}{King}
\address{
  $^1$Stability AI\\
  $^2$University of Edinburgh, UK}
\email{danlyth@gmail.com, Simon.King@ed.ac.uk}

\begin{document}

\maketitle
 
\begin{abstract}
Text-to-speech models trained on large-scale datasets have demonstrated impressive in-context learning capabilities and naturalness. However, control of speaker identity and style in these models typically requires conditioning on reference speech recordings, limiting creative applications. Alternatively, natural language prompting of speaker identity and style has demonstrated promising results and provides an intuitive method of control. However, reliance on human-labeled descriptions prevents scaling to large datasets.

Our work bridges the gap between these two approaches. We propose a scalable method for labeling various aspects of speaker identity, style, and recording conditions. We then apply this method to a 45k hour dataset, which we use to train a speech language model. Furthermore, we propose simple methods for increasing audio fidelity, significantly outperforming recent work despite relying entirely on found data.

Our results demonstrate high-fidelity speech generation in a diverse range of accents, prosodic styles, channel conditions, and acoustic conditions, all accomplished with a single model and intuitive natural language conditioning. Audio samples can be heard at \href{https://text-description-to-speech.com/}{https://
text-description-to-speech.com/}.
\end{abstract}

\section{Introduction}

Scaling both model and training data size has driven rapid progress in generative modeling, especially for text and image synthesis \cite{brown2020language, chowdhery2022palm, ramesh2022hierarchical, rombach2022highresolution}. Natural language conditioning provides an intuitive method for control and creativity in these modalities, enabled by web-scale human-authored text and image annotations \cite{commoncrawl, schuhmann2022laion5b}. However, only recently has speech synthesis started to exploit scale and natural language conditioning.

The initial results from large-scale text-to-speech (TTS) models have demonstrated impressive in-context learning capabilities, such as zero-shot speaker and style adaptation, cross-lingual synthesis, and content editing \cite{wang2023neural, zhang2023speak, le2023voicebox}. However, a reliance on reference speech limits their practical application. It also forces the user to reproduce the likeness of an existing speaker, which is beneficial in some use cases but has the potential for harm, especially when so little enrollment data is required.

To alleviate these shortcomings, the use of natural language to describe speaker and style is starting to be explored (the most recent example is concurrent work to ours, Audiobox. \cite{teamaudiobox2023audiobox}). Unlike the image modality, no large dataset containing natural language descriptions of speech exists, so this metadata must be created from scratch. To date, this has been achieved using a combination of human annotations and statistical measures, with the results often passed through a large language model to mimic the variability that might be expected in genuine human annotations \cite{teamaudiobox2023audiobox, guo2022prompttts, yang2023instructtts, liu2023promptstyle, shimizu2023prompttts}.  However, any approach requiring human annotations is challenging to scale to large datasets. For example, Multilingual LibriSpeech \cite{pratap2020mls} contains over 10 million utterances across 50k hours of audio, equivalent to over five years. Because of the human annotation bottleneck, TTS models using natural language descriptions have been of limited scale and, therefore, unable to demonstrate some of the broad range of capabilities associated with larger models.

In this work, we rely entirely on automatic labeling, enabling us to scale to large data for the first time (along with concurrent work \cite{teamaudiobox2023audiobox}). Coupling this with large-scale speech language models allows us to synthesize speech in a wide range of speaking styles and recording conditions using intuitive natural language control.

Specifically, we:
\begin{enumerate}
\item Propose a method for efficiently labeling a 45k hour dataset with multiple attributes, including gender, accent, speaking rate, pitch, and recording conditions.
\item Train a speech language model on this dataset and demonstrate the ability to control these attributes independently, creating speaker identities and style combinations unseen in the training data.
\item Demonstrate that with as little as 1\% high-fidelity audio in the training data and the use of the latest state-of-the-art audio codec models, it is possible to generate extremely high-fidelity audio.
\end{enumerate}

\begin{figure*}[t!]
    \centering
    \includegraphics[scale=0.26]{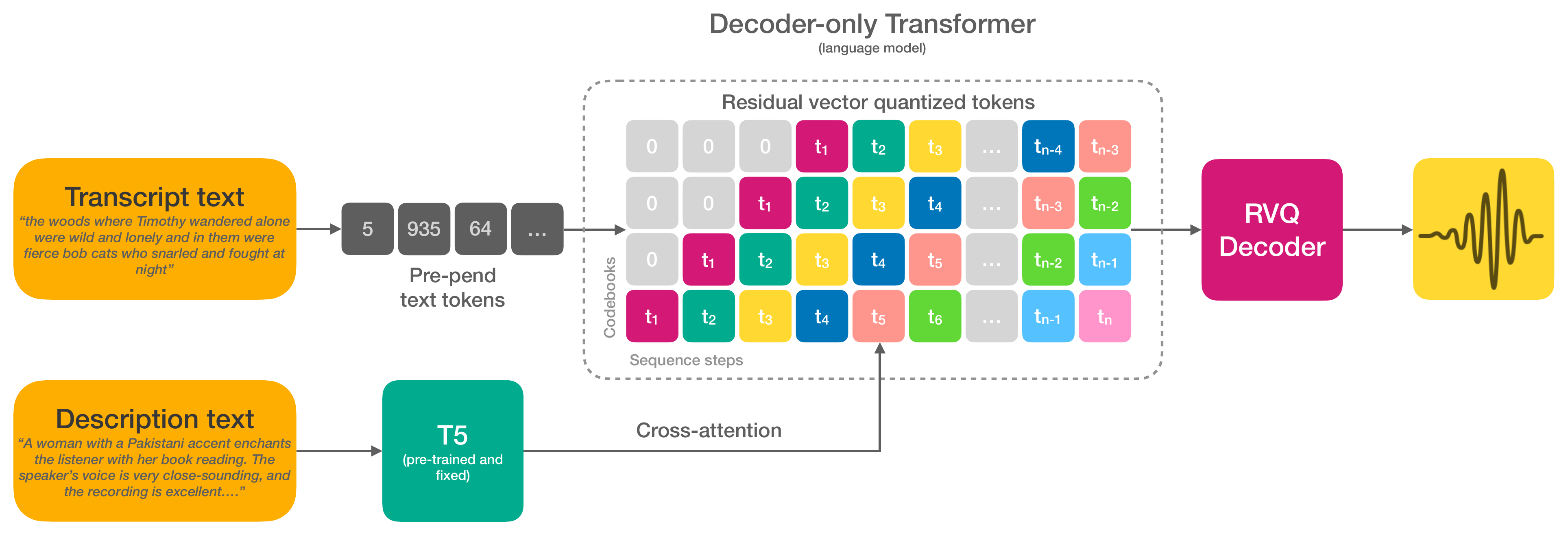}
    \caption{Overview of the model architecture}
    \label{fig:architecture}
\end{figure*}

\section{Related Work}
\subsection{Control of speaker identity and style}

Controlling non-lexical speech information, such as speaking style and speaker identity, has been explored through various approaches. With neural models, the first attempts at this relied on reference embeddings or ``global style tokens'' derived from exemplar recordings \cite{skerry-ryan2018endtoend, wang2018style}. This approach is effective but constrains users to existing recordings, significantly limiting versatility and scalability. To alleviate this, more flexible approaches sample from the continuous latent spaces of Gaussian mixture models and variational autoencoders \cite{hsu2018hierarchical}. However, this approach requires careful training to ensure that the latent variables are disentangled, as well as complex analysis to identify the relationship between these variables and attributes of speech.

In an attempt to bypass the brittleness of reference embeddings and the complexity of disentangled latent variable approaches, recent work has attempted to use natural language descriptions to guide non-lexical speech variation directly. This line of work has no doubt been inspired by the success in other modalities (particularly text-to-image models), but a key challenge for speech is the lack of natural language descriptive metadata.

An initial attempt at circumventing this issue was proposed in \cite{guo2022prompttts}. In this work, the authors use statistically derived metrics (such as speaking rate and pitch) from a dataset of real speech, combined with emotion labels from a dataset of synthetic speech provided by a commercial TTS model. Together, this dataset offers five axes of non-lexical variation, which are turned into keywords, each with three levels of granularity (high, medium, and low).

The authors of \cite{yang2023instructtts} move away from computational labeling methods and explore human annotation. They label 44 hours of data with natural language sentences describing style and emotion but rely on a fixed set of speaker IDs to control speaker likeness. Human annotation is used again in \cite{liu2023promptstyle} (with an even smaller 6-hour dataset) in service of style transfer, and again, speaker likeness is controlled by speaker IDs. However, in \cite{shimizu2023prompttts}, the authors do tackle the labeling and generation of different speaker identities. Human annotations are combined with computational statistics similar to \cite{guo2022prompttts} and are then fed into a language model to create natural language variations. Unlike our work, the authors make no attempt to model accent or channel conditions and only label 16\% of the speakers in a dataset already two orders of magnitude smaller than the one we use. This difficulty in scaling human annotations provides some of the motivation for this work.

While we were running the evaluations for this work, the authors of \cite{teamaudiobox2023audiobox} released Audiobox. This concurrent work uses a similar approach to ours to label a large dataset. However, we propose simple methods to significantly outperform this work in terms of the overall naturalness and audio fidelity of the generated speech. We also outperform this work in how closely our model matches the text description (for those attributes of speech shared across both lines of work). PromptTTS2 \cite{leng2023prompttts2} is also concurrent work that scales to a large dataset, but their approach only attempts to control four variables, significantly limiting the range of capabilities.


\section{Method}\label{method}

\subsection{Metadata collection}\label{metadata_collection}
The data for this study comprises two English speech corpora derived from the LibriVox audiobook project\footnote{\href{https://librivox.org/}{librivox.org}} - the English portion of Multilingual LibriSpeech (MLS) \cite{pratap2020mls} (45k hours) and LibriTTS-R \cite{koizumi2023librittsr} (585 hours). While LibriTTS-R is significantly smaller in scale, we include it given the higher audio quality resulting from enhancement via the Miipher system \cite{koizumi2023miipher}. Both datasets provide transcriptions and a label for gender generated using a predictive model.

\subsubsection{Accent}\label{accent}
Speaker accent is an aspect of speech that natural language prompting in TTS has so far overlooked, despite the wide range of accents found in the datasets typically used.

We appreciate that labeling accents with discrete labels is an ill-formed task considering the discrete-continuous nature of accents. However, the alternative, i.e., ignoring accent altogether, is unacceptable. To this end, we train an accent classifier and use it to label our datasets.

We use EdAcc \cite{sanabria2023edinburgh}, VCTK \cite{yamagishi2012vctk}, and the English-accented subset of VoxPopuli \cite{wang2021voxpopuli} as the training data for our accent classifier. In total, these datasets cover 53 accents. We extract embeddings using the language ID model from \cite{pratap2023scaling} and train a simple linear classifier using these embeddings, achieving an accuracy of 86\% on a held-out test set. We then run this model on our datasets and spot-check the results.

\subsubsection{Recording quality}\label{recording_quality}
Large-scale publicly available speech datasets are typically derived from crowd-sourced projects such as LibriVox. This leads to a fundamental limitation of these datasets - the audio recording quality is often suboptimal compared to professional recordings. For example, many utterances have low signal-to-noise, narrow bandwidth, codec compression artifacts, and excessive reverberation.

To circumvent this limitation, we include LibriTTS-R, a dataset derived from LibriVox but which has, as mentioned, significantly improved audio fidelity. By including this high-fidelity dataset and labeling features related to recording quality across both datasets, we hypothesize that the model will learn a latent representation of audio fidelity. Crucially, this should allow the generation of clean, professional-sounding recordings for accents and styles that only have low-fidelity utterances in the training data.

The two proxies we use for labeling recording quality are the estimated signal-to-noise ratio (SNR) and estimated C50. C50 is the ratio of early reflections to late reflections and indicates how reverberant a recording is. For both of these features, we use the Brouhaha library\footnote{\href{https://github.com/marianne-m/brouhaha-vad}{github.com/marianne-m/brouhaha-vad}} introduced in \cite{lavechin2022brouhaha}.

\subsubsection{Pitch and speaking rate}
We compute pitch contours for all utterances using the PENN library\footnote{\href{https://github.com/interactiveaudiolab/penn}{github.com/interactiveaudiolab/penn}} proposed in \cite{morrison2023crossdomain} and then calculate the speaker-level mean and utterance-level standard deviation. The speaker-level mean is used to generate a label for speaker pitch relative to gender, and the standard deviation is used as a proxy for how monotone or animated an individual utterance is.

The speaking rate is simply calculated by dividing the number of phonemes in the transcript by the utterance length (silences at the beginning and end of the audio files have already been removed). We use the library g2p\footnote{\href{https://github.com/roedoejet/g2p}{github.com/roedoejet/g2p}} for the grapheme-to-phoneme conversion.

\subsection{Metadata preparation}
The next stage is to take all the variables described above and convert them into natural language sentences. To do this, we first create keywords for each variable.

\begin{figure}[t]
  \vskip -6mm
  \centering
  \includegraphics[width=\linewidth]{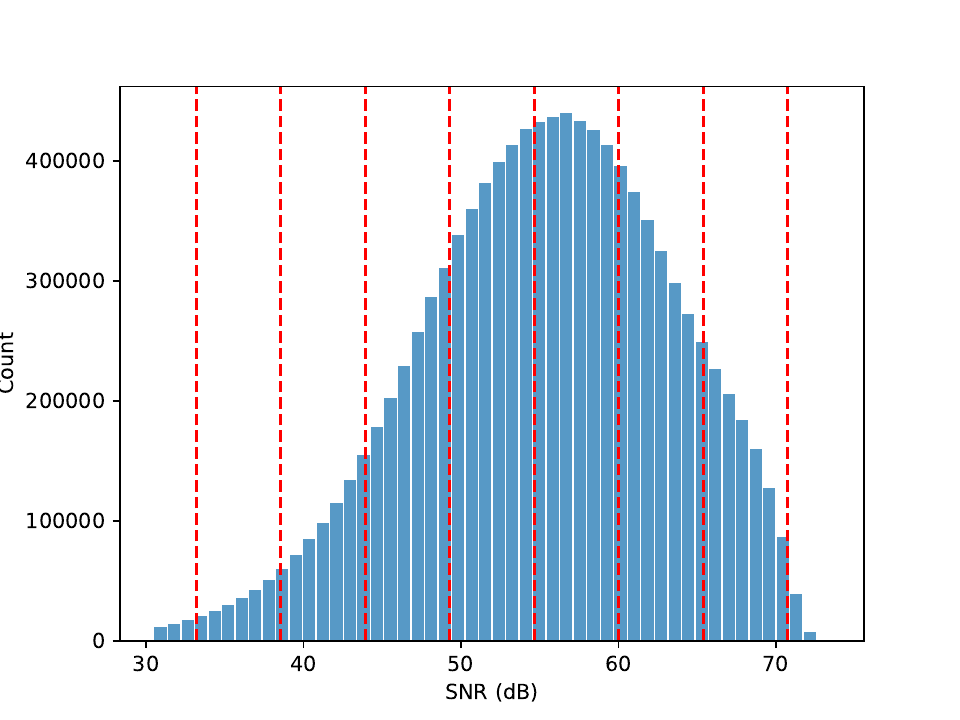}
  \vskip -2mm
  \caption{Estimated SNR across the MLS dataset and the discrete bin boundaries used to create keywords.}
  \label{fig:binning_of_variable}
\end{figure}

The discrete labels such as gender (provided by the dataset creators) and accent require no further processing and can be directly used as keywords. However, the pitch, speaking rate, and estimated SNR and C50 are all continuous variables that must first be mapped to discrete categories. We do this by analyzing the variables across the full dataset and then applying appropriate binning. A visual example of this is shown in Figure \ref{fig:binning_of_variable}, where the estimated SNR across all utterances can be seen along with the bin boundaries. For each variable, we apply seven bins and then use appropriate short phrases to describe each bin. For example, in the case of speaking rate, we use terms such as “very fast”, “quite fast”, “fairly slowly” etc.

Once this binning is complete for all continuous variables, we have keywords for gender, accent, pitch relative to speaker, pitch standard deviation, speaking rate, estimated SNR, and estimated C50. We also create a new category when the SNR and C50 are both in their highest or lowest bin and label these “very good recording” and “very bad recording”, respectively.

To improve generalization and allow the user to input descriptive phrases using their own terminology, we feed these sets of keywords into a language model (Stable Beluga 2.5\footnote{\href{https://huggingface.co/stabilityai/StableBeluga2}{huggingface.co/stabilityai/StableBeluga2}}) with appropriate prompts to create full sentences. For example, “female”, “Hungarian”, “slightly roomy sounding”, “fairly noisy”, “quite monotone”, “fairly low pitch”, “very slowly” could be converted into “a woman with a deep voice speaking slowly and somewhat monotonously with a Hungarian accent in an echoey room with background noise”.

\subsection{Model}\label{model}
We adapt the general-purpose audio generation library AudioCraft\footnote{\href{https://github.com/facebookresearch/audiocraft}{github.com/facebookresearch/audiocraft}} and make it suitable for TTS. This library supports multiple forms of conditioning (text, embeddings, wave files) that can be applied in various ways (pre-pending, cross-attention, summing, interpolation). To date, it has only been used for audio and music generation \cite{kreuk2022audiogen, copet2023simple}. In order to make it suitable for TTS, we remove word drop-out from the transcript conditioning and, after some initial experiments, settle on pre-pending the transcript and using cross-attention for the description (see Figure \ref{fig:architecture}). Unlike previous work, we do not provide any form of audio or speaker embedding conditioning from similar utterances - the model must rely entirely on the text description for gender, style, accent, and channel conditions.

We use the Descript Audio Codec\footnote{\href{https://github.com/descriptinc/descript-audio-codec}{github.com/descriptinc/descript-audio-codec}} (DAC) (44.1kHz version) introduced in \cite{kumar2023highfidelity} to provide our discrete feature representations. This residual vector quantized model produces tokens at a frame rate of 86Hz and has nine codebooks. We use the delay pattern introduced in \cite{copet2023simple} to deal with these nine dimensions in the context of a language model architecture. We choose this codec rather than the popularly used Encodec \cite{defossez2022high} because the authors of DAC demonstrate a subjective and objective improvement in audio fidelity over Encodec. 

\section{Experiments}

\subsection{Objective evaluation}
To evaluate whether our model is capable of generating speech that matches a provided description, we first carry out the following objective evaluations. To test the targeted control of specific attributes, we use the test set sentences from MLS and LibriTTS-R but manually write the description to test the variable of interest. The only exception is the accent test set, where we combine sentences from the Rainbow Passage \cite{fairbanks1960voice}, Comma Gets a Cure \cite{honorof2000comma}, and Please Call Stella. In all cases, we ensure that the descriptions are balanced across the test set.

To test control of gender, we use a pre-trained gender classifier\footnote{\href{https://huggingface.co/alefiury/wav2vec2-large-xlsr-53-gender-recognition-librispeech}{huggingface.co/alefiury/wav2vec2-large-xlsr-53-gender-recognition-librispeech}} that achieves an F1 score of 0.99 on LibriSpeech Clean 100. Using this classifier, our generated test set scores an accuracy of 94\%. In a similar fashion, we re-use our accent classifier (see Section \ref{accent}) and classify the accents of our generated accent test set. Here, we see a somewhat poorer accuracy of 68\%. We hypothesize that this is likely to be due to noisy labeling and a very imbalanced distribution of accents in the training set.

\begin{figure}[t!]
    \centering
    \begin{subfigure}[b]{0.23\textwidth}
        \centering
        \includegraphics[width=\textwidth]{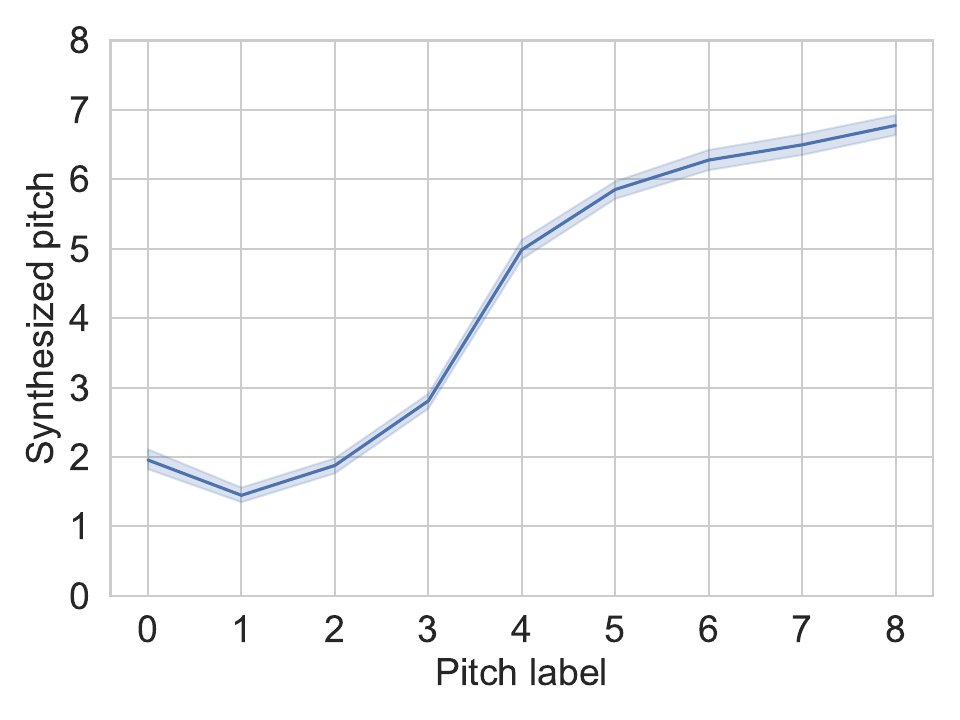}
        \vskip -2mm
        \caption[]%
        {{\small Mean pitch}}    
        \label{fig:mean_pitch}
    \end{subfigure}
    \hfill
    \begin{subfigure}[b]{0.23\textwidth}
        \centering 
        \includegraphics[width=\textwidth]{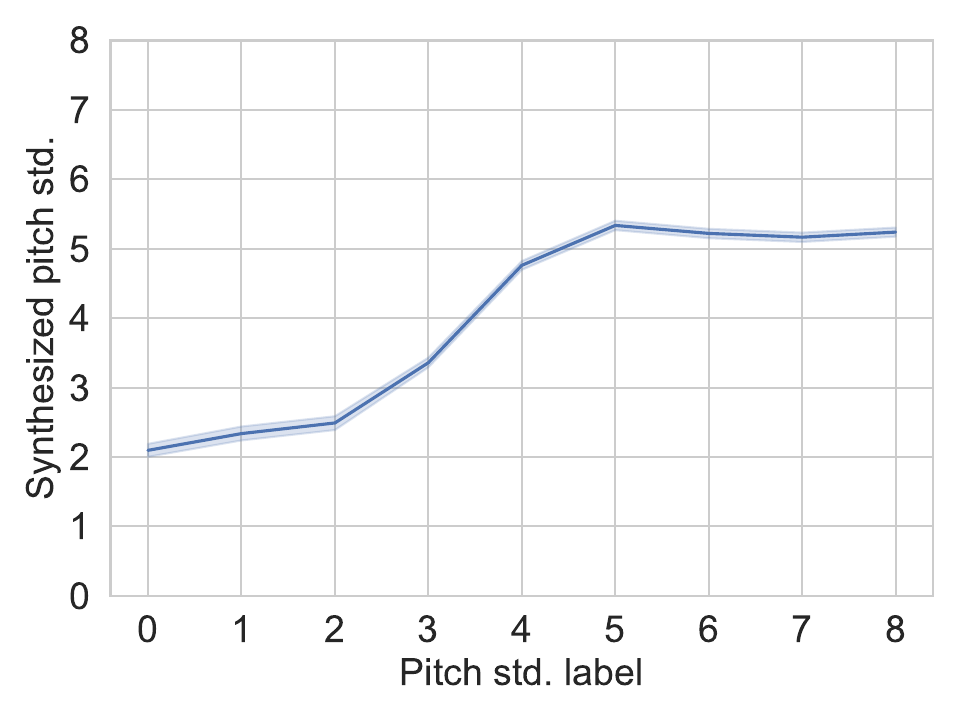}
        \vskip -2mm
        \caption[]%
        {{\small Pitch standard deviation}}    
        \label{fig:pitch_std}
    \end{subfigure}
    \vskip 2mm 
    \begin{subfigure}[b]{0.23\textwidth} 
        \centering 
        \includegraphics[width=\textwidth]{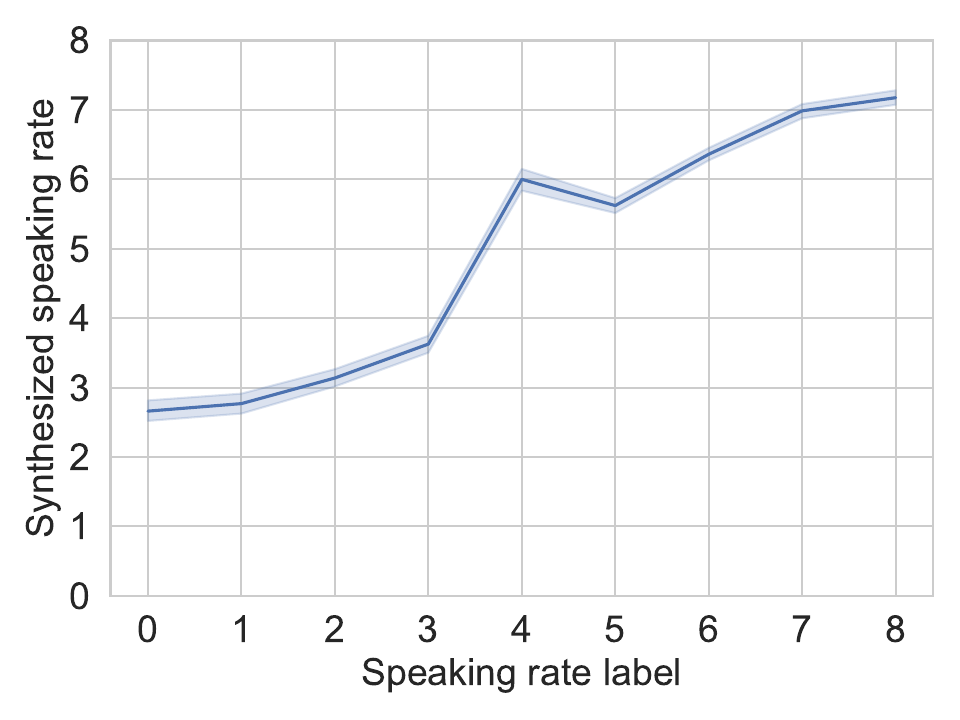}
        \vskip -2mm
        \caption[]%
        {{\small Speaking rate}}    
        \label{fig:speaking_rate}
    \end{subfigure}
    \begin{subfigure}[b]{0.23\textwidth} 
        \centering 
        \includegraphics[width=\textwidth]{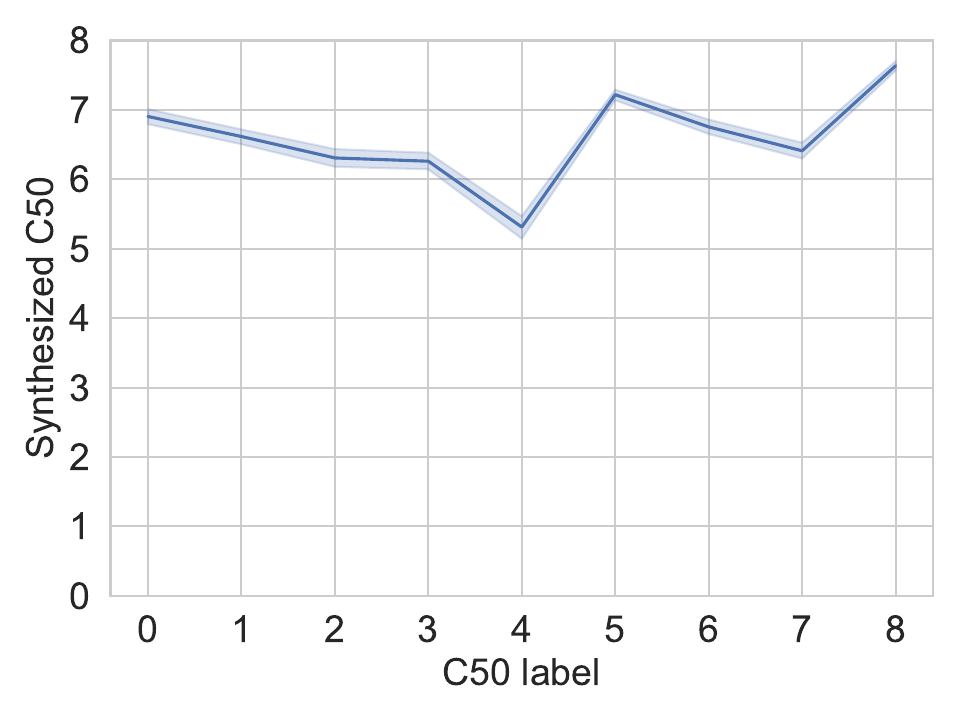}
        \vskip -2mm
        \caption[]%
        {{\small Estimated C50}}    
        \label{fig:c50}
    \end{subfigure}
    \begin{subfigure}[b]{0.23\textwidth} 
        \centering 
        \includegraphics[width=\textwidth]{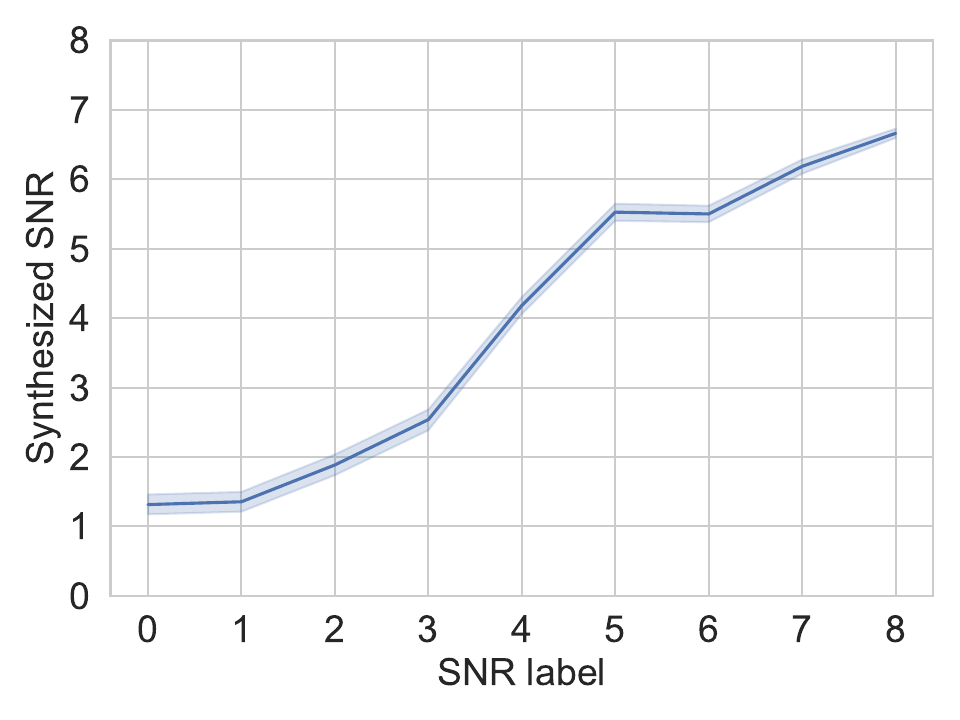}
        \vskip -2mm
        \caption[]%
        {{\small Estimated SNR}}    
        \label{fig:snr}
    \end{subfigure}
    \vskip -2mm
    \caption{Correlation between description labels and synthesized labels (with 95\% confidence intervals).}
    \label{fig:correlation}
\end{figure}

The remaining attributes that we labeled are continuous variables. For these variables, we run our generated test sets through the same models that were used to label the training set (see Section \ref{metadata_collection}). The results can be seen in Figure \ref{fig:correlation}. We see that for every attribute other than C50, the model performs fairly well at generating speech that matches the provided description. We are unsure as to why the model performs poorly at generating audio with the appropriate C50, and further investigation is required.

Our final objective evaluation aims to quantify the audio fidelity of our model when asked to produce audio with “excellent recording quality” or similar terms. Here, we use the recently proposed Torchaudio Speech Quality and Intelligibility Measures \cite{kumar2023torchaudiosquim}. This model provides a reference-less estimate of Wideband Perceptual Estimation of Speech Quality (PESQ),  Short-Time Objective Intelligibility (STOI), and Scale-Invariant Signal-to-Distortion Ratio (SD-SRD). Using 20 test sentences and descriptions from LibriTTS-R, we run these metrics on outputs from our model, Audiobox (using the public website interface\footnote{\href{https://audiobox.metademolab.com/capabilities/tts_description_condition}{audiobox.metademolab.com/capabilities/tts\_description\_condition}}), and the ground truth audio. As can be seen in Table \ref{tab:squim}, our model produces speech with values that are significantly higher than Audiobox and often very close to the ground truth. As mentioned in Section \ref{recording_quality}, this is achieved despite training on only $\sim$500 hours of high-fidelity speech in the context of the full 45k hour dataset.

\subsection{Subjective evaluation}
To complement our objective evaluations, we run two subjective listening tests. The first of these aims to quantify how well our model follows a natural language description (”relevance” or “REL”). An example of such a description taken from the listening test set is: \textit{“A female voice with an Italian accent reads from a book. The recording is very noisy. The speaker reads fairly quickly with a slightly high-pitched and monotone voice.”} We generate 40 sets of samples (20 from MLS and 20 from LibriTTS-R) using descriptions created using the method described in Section \ref{method}. Again, we employ 30 listeners (native English speakers) and ask them to evaluate how closely the speech matches the description using a 5-point scale. For each sentence and description, we present the listeners with outputs from our model, Audiobox, and the ground truth audio. The only processing we apply is loudness normalization to -18 LUFS and the removal of silence before and after speech.

As can be seen in Table \ref{tab:listening_test}, our model outperforms Audiobox in this evaluation. Somewhat counterintuitively, it also outperforms the ground truth. However, there appear to be two clear reasons for this. Firstly, the test set descriptions contain label noise. For example, if a ground truth utterance is labeled with the incorrect accent, the generated speech is likely to be more faithful to the description than the ground truth. Similarly, we also see very occasional instances where samples from LibriTTS-R contain mild audio artifacts (we removed samples containing significant artifacts). In this case, the audio fidelity from our model is likely to be higher than the ground truth and, therefore, closer to the description. One final note on this evaluation is that we are aware that there are aspects of speech that Audiobox is capable of controlling that our model is not (for example, age).

\begin{table}[t!]
  \caption{Speech Quality and Intelligibility Measures (SQUIM) (with 95\% confidence intervals)}
  \label{tab:squim}
  \centering
  \begin{small}
  \begin{tabular}{llll}
    \toprule
    \textbf{Model}      & \textbf{PESQ}         & \textbf{STOI}         & \textbf{SI-SDR}            \\
    \midrule
    Ground truth        & 4.15 \begin{tiny}$\pm{0.04}$\end{tiny}      & 0.997 \begin{tiny}$\pm{0.001}$\end{tiny}      & 27.45 \begin{tiny}$\pm{1.09}$\end{tiny}           \\
    Ours                & 3.84 \begin{tiny}$\pm{0.10}$\end{tiny}      & 0.996 \begin{tiny}$\pm{0.001}$\end{tiny}      & 26.53 \begin{tiny}$\pm{1.16}$\end{tiny}           \\
    Audiobox            & 3.46 \begin{tiny}$\pm{0.16}$\end{tiny}      & 0.988 \begin{tiny}$\pm{0.004}$\end{tiny}      & 21.84 \begin{tiny}$\pm{1.37}$\end{tiny}           \\
    \bottomrule
  \end{tabular}
  \end{small}
\end{table}

\begin{table}[t!]
  \caption{Naturalness and relevance results (with 95\% confidence intervals)}
  \label{tab:listening_test}
  \centering
  \begin{small}
  \begin{tabular}{lll}
    \toprule
    \textbf{Model}      & \textbf{MOS}         & \textbf{REL}                 \\
    \midrule
    Ground truth        & 3.67 \begin{tiny}$\pm{0.09}$\end{tiny}      & 3.62 \begin{tiny}$\pm{0.06}$\end{tiny}          \\
    Ours                & 3.92 \begin{tiny}$\pm{0.07}$\end{tiny}      & 3.88 \begin{tiny}$\pm{0.06}$\end{tiny}          \\
    Audiobox            & 2.79 \begin{tiny}$\pm{0.09}$\end{tiny}      & 3.19 \begin{tiny}$\pm{0.06}$\end{tiny}          \\
    \bottomrule
  \end{tabular}
  \end{small}
\end{table}

Our second listening test aims to evaluate the overall naturalness and audio fidelity of our model. In this case, we only use samples from LibriTTS-R. We ask 30 listeners to rate the speech on a scale of 1-5 (we’ll refer to this simply as mean opinion score, or MOS). We appreciate that this evaluation conflates the naturalness of speech patterns and audio fidelity, but we chose this test design to match that used by the authors of Audiobox. 

In Table \ref{tab:listening_test}, we see that our model significantly outperforms Audiobox. We hypothesize that there are two reasons for this. Firstly, we believe that our use of the DAC codec rather than Encodec has a significant impact (see Section \ref{model} for further details). Secondly,  our use (and labeling) of a small high-fidelity dataset (LibriTTS-R) may well have had a significant impact. However, this hypothesis is difficult to validate without full knowledge of the data used to train Audiobox.

Again, we see our model outperforming the ground truth. As discussed previously, it would appear that this is due to minor audio artifacts in LibriTTS-R caused by the speech-enhancement model. See our demo website\footnote{\href{https://text-description-to-speech.com/}{text-description-to-speech.com}} to hear examples.

\section{Conclusion}
In this work, we propose a simple but highly effective method for generating high-fidelity text-to-speech that can be intuitively guided by natural language descriptions. To the best of our knowledge, this is the first time that such a method is capable of controlling such a wide range of speech and channel condition attributes in conjunction with such high audio fidelity and overall naturalness. In particular, we note that this was possible with an amateurly recorded dataset coupled with a comparatively tiny amount of clean data.

However, this work only demonstrates efficacy in a relatively narrow domain (audiobook reading in English). In the future, we plan to extend to a wider range of languages, speaking styles, vocal effort, and channel conditions.


\bibliographystyle{IEEEtran}
\bibliography{mybib}

\end{document}